\documentclass[a4paper,draft]{article}
\usepackage[
unicode,
pdfauthor={Leron Borsten, Mateo Galdeano, 金炯錄},
pdftitle={Brane Symmetries Revisited: Symmetries of Tensile and Tensionless Branes in Possibly Degenerate Metrics and their Manifestations},
pdfsubject={hep-th,math-ph},
pdflang={en-GB},
breaklinks=true,
final
]{hyperref}
\usepackage[T1,T2A]{fontenc}
\usepackage[unicode,final]{hyperref}
\usepackage{amsmath,amssymb,amsthm,orcidlink,booktabs,mleftright,cleveref,scalerel}
\usepackage[smalltableaux]{ytableau}
\usepackage[noadjust]{cite}
\theoremstyle{definition}
\newtheorem{definition}{Definition}
\usepackage{CJKutf8}
\usepackage[final]{microtype}
\usepackage[latin,russian,british]{babel}
\usepackage[osf]{newpxtext}
\usepackage[varbb,slantedGreek]{newpxmath}
\usepackage{textgreek,mxedruli}
\usepackage[latin,notstar,notbar,notexclam,safe]{armtex}
\renewcommand\armsection\S
\renewcommand\delta\deltaup
\renewcommand\Pi\Piup
\renewcommand\Omega\Omegaup
\renewcommand\varepsilon\varepsilonup
\newcommand\smallcdots{\scaleobj{0.5}{\boldsymbol\cdots}}
\usepackage[utf8]{inputenc}
\newcommand\intprod{\mathbin\lrcorner}

\begin{document}
\title{Brane Symmetries Revisited: Symmetries of Tensile and Tensionless Branes in Possibly Degenerate Metrics and their Manifestations}
\author{Leron Borsten\textsuperscript{\orcidlink{0000-0001-9008-7725}}\and Mateo Galdeano\textsuperscript{\orcidlink{0000-0003-0052-6566}}\and Hyungrok Kim~(\begin{CJK*}{UTF8}{bsmi}金炯錄\end{CJK*})\textsuperscript{\orcidlink{0000-0001-7909-4510}}
\\[1em]
\small{Centre for Mathematics and Theoretical Physics}  \\\small{Department of Physics, Astronomy and Mathematics} \\\small{University of Hertfordshire} \\\small{Hatfield, Hertfordshire\ \textsc{al10 9ab}}\\ \small{United Kingdom}
\\ [1em]
\tt
\{\href{mailto:l.borsten@herts.ac.uk}{l.borsten},\href{mailto:m.galdeano@herts.ac.uk}{m.galdeano},\,\href{mailto:h.kim2@herts.ac.uk}{\texttt{h.kim2}}\}@herts.ac.uk
}
\maketitle
\begin{abstract}
We analyse the symmetries of tensionless and tensile branes moving in a target space with a possibly degenerate metric, with the worldvolume metric remaining nondegenerate. We recover known results about symmetries of strings and branes as well as new results in the tensionless and degenerate-metric cases. We comment on ramifications in the corresponding bulk theories.
\end{abstract}
\tableofcontents

\section{Introduction and summary}
\paragraph{Background.}
Symmetries are an essential tool in the study and understanding of physical theories. Each continuous symmetry gives rise via the Noether theorem to a conserved quantity of the (classical) physical system that remains invariant along the equations of motion. Furthermore, when these symmetries  are free of anomalies  they entail important implications  for the full quantum system. Unfortunately, identifying all the symmetries of a given theory is a complicated process as some of them can be highly non-linear in the derivatives.

An interesting example of this can be observed already for the theory of a particle moving in a gravitational field. It is well-known that this theory is invariant under isometries of the metric \cite{Wald:1984rg}, which are each infinitesimally generated by a Killing vector field $K$ through the transformation
\begin{equation}
    \delta x^\mu=\epsilon K^\mu,
\end{equation}
where $\epsilon$ is the parameter of the transformation. Perhaps somewhat less well-known is that the particle admits additional (non-linear) symmetries parameterised by a generalisation of Killing vector fields known as \emph{Killing tensors} \cite{Walker:1970un}. Every Killing tensor $K$ of rank $k$ defines a symmetry of the theory via \cite{vanHolten:1992bu} 
\begin{equation}
\label{eq:transformation-particle}
    \delta x^\mu=\epsilon K^\mu{}_{\nu_2\cdots\nu_k}\dot x^{\nu_2}\cdots \dot x^{\nu_k}.
\end{equation}
For supersymmetric generalisations of such symmetries, which now correspond to Killing--Yano tensors, see \cite{Rietdijk:1989qa, vanHolten:1992bu, Gibbons:1993ap, Papadopoulos:2007gf, Santillan:2011sh, Papadopoulos:2011cz}. Similarly, extensions to superspace were studied in \cite{Howe:2015bdd, Howe:2016iqw}.
The Killing tensor symmetries  lead to additional conserved quantities, which have been used to study the separability of the Hamilton--Jacobi equations and to identify geodesics in black-hole solutions \cite{Carter:1968rr, Walker:1970un, Woodhouse:1974}, as discussed in numerous  reviews \cite{Penrose:1986ca, Frolov:2008jr, Frolov:2017kze, Lindstrom:2022nrm}.

A similar phenomenon occurs in the context of bosonic non-linear sigma models in string theory, described by the Polyakov action \cite{Brink:1976sc, Deser:1976rb}. Working in lightcone coordinates $(z_+,z_-)$, 
one can define for each Killing tensor $K$ a symmetry transformation known as a $W$-symmetry \cite{Hull:1989wu} (see also \cite{Howe:1991vs, Howe:1991im, Howe:1991ic, deLaOssa:2018iij} for supersymmetric generalisations)
\begin{equation}
\label{eq:transformation-string}
    \delta x^\mu=\epsilon(z_+) K^\mu{}_{\nu_2\cdots\nu_k}\partial_+ x^{\nu_2}\cdots \partial_+ x^{\nu_k} ,
\end{equation}
where now $\epsilon(z_+)$ is allowed to depend on one of the worldsheet coordinates. These chiral symmetries are classical manifestations of the $W$-algebra of extended symmetries of the underlying Conformal Field Theory \cite{Zumino:1979et, Zamolodchikov:1985wn, Hull:1985jv, Banks:1987cy, Odake:1988bh, Eguchi:1988vra, Delius:1989nc, Shatashvili:1994zw, Figueroa-OFarrill:1996tnk, Howe:2006si, Howe:2010az, Alvarez-Consul:2020hbl, Fiset:2021azq, Alvarez-Consul:2023zon, delaOssa:2024cgo}.

Beyond these well understood situations, one can study symmetries generated by Killing tensors for theories of branes. There are two further generalisations of particular interest: allowing degenerate metrics in the target space as well as considering the tensionless limit.

Degenerate metrics are expected to arise in various contexts in general relativity and other gravitational theories. Spacetimes with metrics that have degeneracies have been used to model the initial conditions of the universe \cite{Ellis:1991st,Ellis:1991sp,Klinkhamer:2019frj,Battista:2020lqv,Wang:2021rje,Holdom:2023xip} and in attempts to resolve black hole singularities \cite{Hellaby:1997qm,Capozziello:2024ucm}, as well as in discussions of the path integral of quantum gravity more generally \cite{Tseytlin:1981ks, Jacobson:1987qk, Horowitz:1990qb,Jacobson:1996gu}, cf.\ \cite{Garnier:2025jnp}.
More recently, there has been recent interest in null hypersurfaces (such as black hole horizons), whose induced metrics have degenerate metrics, with close links to Carrollian and Galilean symmetries \cite{Duval:2014uva,Harmark:2017rpg,Donnay:2019jiz} (see the reviews \cite{Bergshoeff:2022eog,Hartong:2022lsy,Oling:2022fft}).

Tensionless limits of branes are another interesting `degenerate' limit of the standard \(p\)-brane action.
For \(p=0\), this corresponds to the massless limit of the particle, for which the worldline action's symmetry enhances from Killing tensors to conformal Killing tensors \cite{Walker:1970un}. For the string, this is the tensionless string \cite{Schild:1976vq,Karlhede:1986wb,Zheltukhin:1988rys,Zheltukhin:1989ar,Barcelos-Neto:1989pjq,Gamboa:1989zc,Lindstrom:1990ar,Lindstrom:1990qb,Isberg:1993av}, closely related to Galilean symmetries \cite{Bagchi:2013bga,Bagchi:2015nca}, which again enjoys enhanced symmetries \cite{Lindstrom:2022iec}, and may be more feasible for quantisation than tensile branes for \(p>1\) \cite{Dutta:2024gkc}. A tensionless limit for arbitrary bosonic D-$p$-branes was studied in \cite{Lindstrom:1997uj}.

\paragraph{Results.}
We study the bosonic non-linear sigma-model describing a brane with \((p+1)\)-dimensional worldvolume moving through spacetime in the presence of a gauge field and a dilaton \cite{Hughes:1986fa, Bergshoeff:1987cm, Achucarro:1987nc, Bergshoeff:1987qx}. Therefore, we assume the perspective of the $(p+1)$-dimensional worldvolume theory (usually called worldline for $p=0$ and worldsheet for $p=1$), and explore the symmetries of a Polyakov-like action under a generalisation of the transformations \eqref{eq:transformation-particle} and \eqref{eq:transformation-string}. We allow for the target-space metric to be degenerate along the directions normal to the brane (but not parallel to the brane since we assume a nondegenerate worldvolume metric), and we also explore the limit where the brane becomes tensionless (due to the dilaton diverging in a particular way).

For the tensile brane moving in a target space with a nondegenerate metric, we recover the known symmetries. However, more interesting symmetries appear in the tensionless limit or in the presence of a degenerate target space. For the tensionless string without a background gauge field, i.e.\ Kalb--Ramond field, the symmetries found were previously discussed in \cite{Lindstrom:2022iec}. For the zero-brane, i.e.\ a particle, the `dilaton' \(\phi\) appears as a position-dependent mass; the analysis then yields the symmetries of a charged particle with position-dependent mass coupled to a background Maxwell field.

As a byproduct, in the case with a nondegenerate metric, we show that the \(p\)-brane Hamiltonian may be  explicitly written in terms of the Duff--Lu generalised metric for \(p\)-branes \cite{Duff:1989tf,Duff:1990hn} (see \cref{ssec:hamiltonian-duff-lu})  that manifests U-duality\footnote{Here we are using U-duality to mean the na\"ive expected  duality groups for a generic $p$-brane, as described in \cite{Duff:1990hn}, which matches the U-dualities of M-theory proper for M$2$-branes in  target space of dimension $d\leq 4$.}; this generalises the Hamiltonian for M2-branes given in \cite{Strickland-Constable:2021afa,Hatsuda:2012vm}.

Furthermore, we discuss the ways in which the symmetries of brane worldvolume actions appear in other contexts. As mentioned above, in the string case they provide a classical limit of the underlying quantum $W$-algebra on the worldsheet (see \cite{delaOssa:2024cgo} for a recent discussion on the topic). We explain how a quantisation procedure could be performed in the general case and work towards the construction of a worldvolume analogue of the worldsheet $W$-algebras. Similarly, the Noether currents of the brane worldvolume symmetries may be construed as invariants of trajectories of the classical branes moving through spacetime, which may be thought of as generalised geodesics \cite{Strickland-Constable:2021afa}; from this perspective our results generalise the existing discussions for the particle case (where the generalised geodesics are true geodesics, giving rise to conserved quantities along them \cite{Wald:1984rg, Carter:1968rr, Walker:1970un}) as well as for strings in \cite{Chervonyi:2013eja, Chervonyi:2015ima}.

\paragraph{Future directions.}
In this paper, we limit ourselves to purely bosonic branes. We expect similar results for fermions, and in particular for the supersymmetric brane actions, in the spirit of \cite{Rietdijk:1989qa, vanHolten:1992bu, Gibbons:1993ap, Papadopoulos:2007gf, Santillan:2011sh, Papadopoulos:2011cz, Howe:1991vs, Howe:1991im, Howe:1991ic, deLaOssa:2018iij}, but phrased in terms of Killing--Yano tensors rather than Killing tensors.

There are certain similarities between Killing tensors, which are totally symmetric tensors, and higher-form symmetries with differential-form currents, which are totally antisymmetric tensors, as introduced in the seminal work \cite{Gaiotto:2014kfa} and reviewed in \cite{Cordova:2022ruw,Brennan:2023mmt,Gomes:2023ahz,Bhardwaj:2023kri,Luo:2023ive,Schafer-Nameki:2023jdn}. It is known that gravitational theories enjoy many interesting generalised notions of symmetries \cite{Hinterbichler:2022agn, Gomez-Fayren:2023qly,Hull:2024bcl,Hull:2024ism,Cheung:2024ypq,Caro-Perez:2024akv,Hull:2025ivk,Apruzzi:2025hvs,Borsten:2025pvq}; it would be interesting to explore how or whether brane worldvolume symmetries discussed here fit in this landscape.

\paragraph{Organisation of this paper.}
The structure of the paper is as follows: in \Cref{sec:review-of-tensors} we remind the reader of the definition of Killing and conformal Killing tensors. In \cref{sec:warmup} we quickly review how Killing tensors arise as symmetries of the Polyakov particle action and their interpretation as geodesic or bulk symmetries. \Cref{sec:brane-symmetries} is the core of the paper, where we introduce our generalisation of $W$-symmetries and show how it particularises to the cases of a particle, a string and a 2-brane. We conclude commenting on different applications of our symmetry in \Cref{sec:applications}: the geodesic interpretation is discussed in \Cref{sec:noether-charge}, the potential to obtain analogues of $W$-algebras is explored in \Cref{sec:w-algebra} and we also highlight the bulk perspective in \Cref{sec:conserved-quantities}.

\paragraph{Conventions.}
We use the mostly-plus metric signature. The d'Alembertian of a pseudo-Riemannian manifold is \(\square\coloneqq g^{\mu\nu}\nabla_\mu\nabla_\nu\). Index (anti-)symmetrisations are normalised, e.g.\ \(T_{[\mu\nu]}=\frac12(T_{\mu\nu}-T_{\nu\mu})\).

Worldsheet Lorentz indices are denoted \(\alpha,\beta,\dotsc\) while target-space (bulk) Lorentz indices are denoted \(\mu,\nu,\dotsc\). The notation \(\intprod\) denotes the interior product between a vector field and a differential form.

\section{Lightning review of Killing tensors}
\label{sec:review-of-tensors}
We briefly recall the definition of Killing tensors. For reviews, see  \cite{Santillan:2011sh,Cariglia:2014ysa,Frolov:2017kze,Lindstrom:2022nrm}.

Let \(M\) be a \(d\)-dimensional smooth manifold equipped with an affine connection \(\nabla\).
\begin{definition}\label{def:killing-tensor}
A \emph{Killing tensor} (or sometimes a \emph{Killing--Stäckel tensor}) of rank \(k\) is a totally symmetric tensor \(K_{\mu_1\dotso\mu_k}=K_{(\mu_1\dotso\mu_k)}\) such that the following holds:
\begin{equation}\label{eq:killing-tensor-definition}
    \nabla_{(\mu_0}K_{\mu_1\dotso\mu_k)}=0.
\end{equation}
\end{definition}
That is, in representation-theoretic terms, the covariant derivative of a totally symmetric tensor of rank \(k\) transforms under the \(\operatorname{GL}(d)\)-representation
\begin{equation}
    \ydiagram1 \otimes \overbrace{\ytableaushort{{}{}{\none[\smallcdots]}{}}}^k
    =
    \overbrace{\ytableaushort{{}{}{}{\none[\smallcdots]}{}}}^{k+1}
    \oplus
    \overbrace{\ytableaushort{{}{}{\none[\smallcdots]}{},{}}}^k,
\end{equation}
and the Killing-tensor condition requires that the first irreducible component vanish. Note that \eqref{eq:killing-tensor-definition} is precisely the commuting-indices analogue of the condition for a closed differential \(p\)-form \(\nabla_{[\mu_0}A_{\mu_1\dotso\mu_p]}=0\), so that Killing tensors may be regarded in a certain sense as odd analogues of higher-form symmetries whose currents are closed differential forms. Further note that a covariantly constant symmetric tensor is always a Killing tensor (just like how a covariantly constant differential form is always closed).

Suppose further that the connection \(\nabla\) is the Levi-Civita connection of a (pseudo-)Riemannian metric \(g\) of signature \((p,d-p)\). Then we may make a finer decomposition into irreducible \(\operatorname{SO}(p,d-p)\)-representations, which enables the following definition.
\begin{definition}
A \emph{conformal Killing tensor} of rank \(k\) is a totally symmetric traceless tensor \(K_{\mu_1\dotso\mu_k}=K_{(\mu_1\dotso\mu_k)}\) such that the following holds:
\begin{equation}
    \nabla_{(\mu_0}K_{\mu_1\dotso\mu_k)}=
    g_{(\mu_0\mu_1}K'_{\mu_2\dotso\mu_k)}
\end{equation}
for some totally symmetric tensor \(K'_{\mu_2\dotso\mu_k}\).
\end{definition}
It follows from the definition that
\begin{equation}
    K'_{\mu_2\dotso\mu_k}=
    \frac k{d+2(k-1)}\nabla^{\mu_1}K_{\mu_1\mu_2\dotso\mu_k}.
\end{equation}
That is, in representation-theoretic terms, the covariant derivative of a totally symmetric tensor of rank \(k\) transforms under the \(\operatorname{SO}(p,d-p)\)-representation
\begin{equation}
    \ydiagram1 \otimes \overbrace{\ytableaushort{{}{}{\none[\smallcdots]}{}}}^k
    =
    \overbrace{\ytableaushort{{}{}{}{\none[\smallcdots]}{}}}^{k+1}
    \oplus
    \overbrace{\ytableaushort{{}{}{\none[\smallcdots]}{},{}}}^k
    \oplus
    \overbrace{\ytableaushort{{}{\none[\smallcdots]}{}}}^{k-1}
\end{equation}
(since we are now considering \(\operatorname{SO}(p,d-p)\) representations, \(\ytableaushort{{}{\none[\smallcdots]}{}}\) now represents a totally traceless symmetric tensor rather than a mere symmetric tensor), and for a conformal Killing tensor we require that
the first term vanish. While (confusingly) not every Killing tensor is a conformal Killing tensor (because it may not be traceless), every Killing tensor of rank \(k\) may be decomposed into a sum of conformal Killing tensors of ranks \(k\), \(k-2\), \(k-4\), etc.\ by subtracting traces.

\subsection{Killing tensors with a non\texorpdfstring{-}{‐}invertible metric}\label{ssec:degenerate_killing}
With an invertible metric \(g_{\mu\nu}\), note that, given a \((1,k-1)\)-tensor \(K\) such that \(g_{\mu_1\nu}K^\nu{}_{\mu_2\dotso\mu_k}\) is totally symmetric, the coordinate expression
\begin{equation}\label{eq:degenerate_covariant_derivative}
\begin{aligned}
    \nabla_{(\mu_0}K_{\mu_1\dotso\mu_k)}
    &=
    \partial_{(\mu_0}K_{\mu_1\dotso\mu_k)}
    -
    \Gamma^\nu_{(\mu_0\mu_1|}K_{\nu|\mu_2\dotso\mu_k)}
    -
    \dotsb
    -
    \Gamma^\nu_{(\mu_0\mu_k}K_{\mu_1\dotso\mu_{k-1})\nu}\\
    &=
    \partial_{(\mu_0}(g_{\mu_1|\rho}K^\rho{}_{|\mu_2\dotso\mu_k)})
    -
    k\Gamma_{\nu(\mu_0\mu_1}K^\nu{}_{\mu_2\dotso\mu_k)}
\end{aligned}
\end{equation}
can entirely be written in terms of Christoffel symbols of the first kind \(\Gamma_{\mu\nu\rho}\). This enables us to extend the definition of Killing tensors even when \(g\) is degenerate as follows.

Suppose that \(M\) is a smooth manifold and \(g\) is a not necessarily invertible symmetric \((0,2)\)-tensor field on \(M\). The non-invertibility of \(g\) means that it does not always define an affine connection on \(M\). Concretely, while Christoffel symbols of the second kind (the components of the Levi-Civita connection)
\begin{equation}
    \Gamma^\mu_{\nu\rho}
    =\frac12g^{\mu\sigma}
    \left(
        \partial_\nu g_{\sigma\rho}
        +\partial_\rho g_{\sigma\nu}
        -\partial_\sigma g_{\nu\rho}
    \right)
\end{equation}
may be undefined due to the presence of the inverse \(g^{\mu\sigma}\), Christoffel symbols of the first kind
\begin{equation}
    \Gamma_{\mu\nu\rho}
    =\frac12\left(
        \partial_\nu g_{\mu\rho}
        +\partial_\rho g_{\mu\nu}
        -\partial_\mu g_{\nu\rho}
    \right)
\end{equation}
are well defined even for non-invertible \(g\).
In this case, expressions such as \eqref{eq:killing-tensor-definition} do not make sense.

However, the last line of \eqref{eq:degenerate_covariant_derivative} continues to makes sense for degenerate $g$, so that we can define a Killing tensor in this case to be a tensor \(K\) of rank \((1,k-1)\) such that \(g_{\mu_1\nu}K^\nu{}_{\mu_2\dotso\mu_k}\) is totally symmetric and such that
\begin{equation}
    \partial_{(\mu_0}g_{\mu_1|\rho}K^\rho{}_{|\mu_2\dotso\mu_k)}
    -
    k\Gamma_{\nu(\mu_0\mu_1}K^\nu{}_{\mu_2\dotso\mu_k)} = 0.
\end{equation}
Conformal Killing tensors are defined similarly in the degenerate case. In the case \(k=1\), this then reduces to the definition of a (conformal) Killing vector \(\mathcal L_Kg=0\) (or \(\mathcal L_Kg\propto g\)), which makes sense regardless of the nondegeneracy of \(g\).
Similarly, (conformal) Killing--Yano tensors (see \cite{Lindstrom:2022nrm}) make sense with respect to a degenerate metric as long as one has a tensor of rank \((1,k-1)\) rather than of \((0,k)\).

\section{Warm\texorpdfstring{-}{‐}up: Killing tensors from particle Polyakov action}\label{sec:warmup}
As a warm-up exercise, we consider the symmetries of a free particle moving in curved spacetime from the worldline, geodesic, and bulk perspectives and show that the symmetries correspond to Killing tensors (for the massive case) or conformal Killing tensors (for the massless case).

\paragraph{Killing tensors from the Polyakov action.}
Consider a scalar particle of mass \(m\) living in a spacetime \(M\) with invertible (pseudo-)Riemannian metric \(g\).
It is described by the worldline Polyakov action \cite{Brink:1976sz, Brink:1976uf}
\begin{equation}\label{eq:particle-polyakov-action}
    S_\mathrm{particle}[x,e]=\int_\Sigma\mathrm dt\,\frac12\left(e^{-1}g_{\mu\nu}(x)\dot x^\mu\dot x^\nu-em^2\right),
\end{equation}
where \(\Sigma\) is the worldline (parameterised by a real number \(t\)), \(x\colon\Sigma\to M\) is the embedding map, \(e\in\Gammaup(\mathrm T^*\Sigma)\) is the einbein on \(\Sigma\), and \(\dot x\) is the derivative of \(x\) with respect to \(t\).
Assuming \(e\) to be positive everywhere, we may gauge-fix the reparameterisation invariance by going to the gauge \(e=1\), so that the action reduces to
\begin{equation}\label{eq:gauge-fixed-particle-polyakov-action}
    S[x]=\int_\Sigma\mathrm dt\,\frac12\left(g_{\mu\nu}(x)\dot x^\mu\dot x^\nu-m^2\right)
\end{equation}
at the cost of having to impose by hand the on-shell constraint
\begin{equation}\label{eq:on-shell-constraint}
    \dot x^\mu\dot x^\nu g_{\mu\nu}(x)=-m^2
\end{equation}
that results from the equation of motion for \(e\).
Under a general variation of \(x\), the action varies as
\begin{equation}
    \delta S =-\int _\Sigma\mathrm dt\,\delta x^\mu\left(\Gamma(x)_{\mu\nu\rho}\dot x^\nu\dot x^\rho+g(x)_{\mu\nu}\ddot x^\nu\right),
\end{equation}
where \(\Gamma_{\mu\nu\rho}=\frac12(\partial_\nu g_{\mu\rho}+\partial_\mu g_{\nu\rho}-\partial_\mu g_{\nu\rho})\) are the Christoffel symbols of the first kind for \(g\), so that the equation of motion for \(x\) is the geodesic equation
\begin{equation}\label{eq:particle-geodesic-equation}
    \dot x^\mu\nabla_\mu\dot x^\nu=0
\end{equation}
on the (pseudo-)Riemannian manifold \(M\), where we have used $\partial_t = \dot x^\mu \partial_\mu$.

Let \(K\) be a totally symmetric tensor field of rank \(k\) on \(M\). Consider the variation
\begin{equation}\label{eq:particle-howe-papadopoulos-variation}
    \delta x^\mu(t)= \epsilon(t)\dot x^{\nu_2}(t)\dotsm\dot x^{\nu_k}(t)K^\mu{}_{\nu_2\dotso\nu_k}(x(t)),
\end{equation}
where \(\epsilon(t)\) is the symmetry parameter. The on-shell constraint \eqref{eq:on-shell-constraint} then requires that
\begin{equation}
    0=g_{\mu\nu}\dot x^\mu(t)\,\delta\dot x^\nu(t)
    +\frac12\dot x^\mu(t)\dot x^\nu(t)\delta x^\rho\partial_\rho g_{\mu\nu},
\end{equation}
which is inconsistent with the total symmetry of \(K\).
However, the constraint \eqref{eq:on-shell-constraint} is a first-class constraint in that the equation of motion, namely the geodesic equation \eqref{eq:particle-geodesic-equation}, guarantees that the invariant squared mass \(-g_{\mu\nu}\dot x^\mu\dot x^\nu\) never changes:
\begin{equation}
    \dot x^\rho\nabla_\rho(-g_{\mu\nu}\dot x^\mu\dot x^\nu)=0.
\end{equation}
Thus, we may consistently drop the constraint \eqref{eq:on-shell-constraint}.
The theory then describes a particle of unspecified but constant mass given by \(-g(x)_{\mu\nu}\dot x^\mu\dot x^\nu\), which is now the  conserved quantity corresponding to the Killing tensor \(g\).

Integrating by parts, the variation of the action \eqref{eq:gauge-fixed-particle-polyakov-action} under \eqref{eq:particle-howe-papadopoulos-variation} is
\begin{equation}
    \delta S = \frac12\int_\Sigma\mathrm dt\,
    \left(\dot\epsilon(t)K_{\mu_1\dotso\mu_k}\dot x^{\mu_1}\dotsm\dot x^{\mu_k}+\epsilon(t)(\nabla_{\mu_0}K_{\mu_1\dotso\mu_k})
    \dot x^{\mu_0}\dotsm\dot x^{\mu_k}\right).
\end{equation}
Requiring the first term to vanish means that \(\epsilon\) must be constant. Then the remaining necessary and sufficient condition for \(\delta S=0\) for arbitrary \(x\) and constant \(\epsilon\) is therefore that \(K\) must be a Killing tensor:
\begin{equation}\label{eq:particle-cond-2a}
    \nabla_{(\mu_0}K_{\mu_1\dotso\mu_k)}=0.
\end{equation}
In the massless case, since we have the constraint \(\dot x^2=0\), we may relax \eqref{eq:particle-cond-2a} to 
\begin{equation}\label{eq:particle-cond-2-massless}
    \nabla_{(\mu_0}K_{\mu_1\dotso\mu_k)}=g_{(\mu_0\mu_1}\tilde K_{\mu_2\dotso\mu_k)}
\end{equation}
for some totally symmetric tensor \(\tilde K_{\mu_2\dotso\mu_k}\), i.e.\ \(K\) is then a conformal Killing tensor.

\paragraph{Killing tensors from geodesics.}
The Noether charge \(Q_K\) corresponding to the transformation \eqref{eq:particle-howe-papadopoulos-variation}, namely
\begin{equation}\label{eq:particle-conserved-quantity}
    Q_K=K_{\mu_1\dotso\mu_k}\dot x^{\mu_1}\dotsm\dot x^{\mu_k},
\end{equation}
is conserved along geodesics. This fact is well-known \cite{Walker:1970un} (reviewed in \cite{Penrose:1986ca} and \cite[§2.1]{Lindstrom:2022nrm}) and can be seen directly from the geodesic equation
\begin{equation}
    \dot x^\mu\nabla_\mu\dot x^\nu = 0.
\end{equation}
That is, one can directly compute for a Killing tensor \(K\)
\begin{multline}\label{eq:killing_geodesic_particle}
    \dot x^\mu\nabla_\mu(K_{\nu_1\dotso\nu_k}\dot x^{\nu_1}\dotsm\dot x^{\nu_k})=\\
    \dot x^\mu(\nabla_\mu K_{\nu_1\dotso\nu_k})\dot x^{\nu_1}\dotsm\dot x^{\nu_k}+
    kK_{\nu_1\dotso\nu_k}\dot x^{\nu_1}\dotsm\dot x^{\nu_{k-1}}(\dot x^\mu \nabla_\mu\dot x^{\nu_k})= 0.
\end{multline}
Similarly, when \(K\) is a conformal Killing tensor and \(x(t)\) is a null geodesic (the trajectory of a massless particle) we also find
\begin{equation}\label{eq:conformal_killing_geodesic_particle}
    \dot x^\mu\nabla_\mu(K_{\nu_1\dotso\nu_k}\dot x^{\nu_1}\dotsm\dot x^{\nu_k})=0.
\end{equation}

\paragraph{Killing tensors from the bulk theory.}
The particle Polyakov action may be second-quantised to yield a free scalar theory on \(M\), namely
\begin{equation}
    S = \frac12\int_M\mathrm d^{\dim M}x\,\sqrt{|\det g|}\phi(\square-m^2)\phi,
\end{equation}
where the d'Alembertian is \(\square=g^{\mu\nu}\nabla_\mu\nabla_\nu\).
Consider the space of solutions to the corresponding (Klein--Gordon) equation of motion:
\begin{equation}
    \mathcal C=\{\phi\in\mathcal C^\infty(M,\mathbb R)|(\square-m^2)\phi=0\}.
\end{equation}
Consider differential operators of order \(k\) that act on \(\phi\):
\begin{equation}
    T = T_{\mu_1\dotso\mu_k}\nabla_{\mu_1}\dotsm\nabla_{\mu_k}+\text{lower-order terms}.
\end{equation}
The operator \(T\) defines a symmetry\footnote{This follows the terminology of \cite{Eastwood:2002su}.} of the space of solutions \(\mathcal C\) of the free scalar theory if \(T\mathcal C\subseteq\mathcal C\).
It is easy to see that, for \(m\ne0\), this happens when \(T\) commutes with \(\square\); for \(m=0\), we may relax this so that \(T\) is a symmetry of \(\mathcal C\) if there exists another differential operator \(T'\) so that
\begin{equation}
    \square T = T'\square.
\end{equation}
For \(m\ne0\), symmetries are in bijection with Killing tensors \cite{mss}: \(T\) is a symmetry if and only if the leading-order coefficient \(T_{\mu_1\dotso\mu_k}\) corresponds to a Killing tensor \cite{mss}, and conversely every Killing tensor determines the leading-order coefficient of a unique symmetry \(T\).
For \(m=0\), symmetries are in bijection with conformal Killing tensors up to equivalence\footnote{
Here, two symmetries \(T_1\) and \(T_2\) are equivalent if and only if \(T_1-T_2=T'\square\) for some differential \(T'\). Clearly, every differential operator of the form \(T'\square\) is a symmetry of the massless scalar field theory, so such trivial symmetries should be quotiented out.}  \cite{Eastwood:2002su}:
\(T\) is a symmetry if and only if it is equivalent to a differential operator whose leading-order coefficient \(T_{\mu_1\dotso\mu_k}\) corresponds to a conformal Killing tensor, and conversely every conformal Killing tensor determines the leading-order coefficient of a unique symmetry \(T\) up to equivalence.

A direct correspondence between the two sets of symmetries is given by passing through the worldline formalism for quantum field theory \cite{Corradini:2015tik,Schubert:2001he,Schubert:1996jj,Schmidt:1994da}.
Concretely, the propagator formally takes the form
\begin{align}
    G(x_0,x_1) &=
    \int_{\operatorname{Einbein}(\Sigma)}\mathrm De\int_{\operatorname{Path}(x_0,x_1)}\mathrm Dx\,\exp\left(\mathrm iS_\mathrm{particle}\right)\\
    &=
    \int_0^\infty\mathrm dL \int_{\operatorname{Path}(x_0,x_1)}\mathrm Dx\,\exp\left(\frac{\mathrm i}2\int_0^L\mathrm dt\,(\dot x^2-m^2)\right),
\end{align}
where \(S_\mathrm{particle}\) is the particle Polyakov action \eqref{eq:particle-polyakov-action}, \(\operatorname{Path}(x_0,x_1)\) is the space of paths from \(x_0\) to \(x_1\), and \(\operatorname{Einbein}(\Sigma)\) is the space of einbeins on \(\Sigma\); after gauge-fixing, the integral over \(e\) reduces to the invariant length \(L\) of \(\Sigma=[0,1]\), and an \(N\)-point correlation function of the free scalar field theory is a sum of products of propagators \(G(x_0,x_1)\). Insofar as \(G(x_0,x_1)\) is defined in terms of the particle worldline action \(S_\mathrm{particle}\) \eqref{eq:particle-polyakov-action}, it directly inherits symmetries of \(S_\mathrm{particle}\), and so does the \(N\)-point correlation function of the second-quantised field theory.

\section{Symmetries of bosonic branes}
\label{sec:brane-symmetries}
\subsection{The gauge\texorpdfstring{-}{‐}fixed \texorpdfstring{\(p\)-}{𝑝‐}brane action}
Consider the \(p\)-brane with worldvolume \(\Sigma\) moving inside a \(d\)-dimensional spacetime manifold \(M\) with a possibly degenerate string-frame metric\footnote{That is, a symmetric tensor of rank \((0,2)\); later, \eqref{eq:worldvolume-metric-eom} implies that the brane is then stretched along nondegenerate directions, but transverse directions may be degenerate} \(\tilde g\)
and coupled to a \((p+1)\)-form potential \(A\) as well as a dilaton \(\phi\).
An action is given by\footnote{This factor is correct for a D-brane; for other branes one may need to introduce a constant factor in front of \(\phi\). For instance, for an NS5-brane, the factor should be \(\exp(-2\phi)\). This does not affect the symmetry analysis.}
\begin{equation}\label{eq:brane-action-orig}
    S_p[\gamma,x] = \int_\Sigma\mathrm d^{p+1}z\,\frac12 \sqrt{|\det\gamma|}\exp(-\phi)\left(-\tilde g_{\mu \nu}\gamma^{\alpha \beta}\partial_\alpha x^\mu \partial_\beta x^\nu+p-1\right)+x^*A,
\end{equation}
where \(x\colon\Sigma\to M\) is the embedding map of the brane and \(\gamma\) is a nondegenerate\footnote{While the target-space metric may be degenerate, we assume the worldvolume metric \(\gamma\) to be nondegenerate in order to have a well-defined action.} pseudo-Riemannian metric on the worldsheet, and \(x^*A\) is the pullback of the \((p+1)\)-form \(A\in\Omega^{p+1}(M)\), which we assume for simplicity to be defined on \(M\) globally,\footnote{
    Of course, one can more generally have situations where the connection \(A\) is only defined patchwise and not globally. In this case, global issues may break the continuous symmetries discussed here into certain discrete subgroups \cite{Arias-Tamargo:2025xdd}. We neglect this issue in what follows.
} onto the worldsheet via \(x\).
This is the action given in e.g.\ \cite[(7.1)]{Duff:1990hn} with an additional coupling to the dilaton \(\phi\) \cite{Johnson:2003cvf}.
For our purposes, it is convenient to define \(\tilde\Phi(x)\coloneqq\exp(-\phi(x)/2)\) and pass to the Einstein-frame metric \(g(x)\coloneqq\tilde\Phi(x)^2\tilde g(x)\) to write
\begin{equation}\label{eq:brane-action}
    S_p[\gamma,x] = \int_\Sigma\mathrm d^{p+1}z\,\frac12 \sqrt{|\det\gamma|}\left(-g_{\mu\nu}\gamma^{\alpha \beta}\partial_\alpha x^\mu \partial_\beta x^\nu+(p-1)\tilde\Phi(x)^2\right)+x^*A,
\end{equation}
such that, when \(p\ne1\), \(\tilde\Phi\) gives a position-dependent worldvolume cosmological constant, i.e.\ tension (or, for \(p=0\), position-dependent particle mass).

This action enjoys worldvolume diffeomorphism symmetry. (Additionally, for \(p=1\), the action is also invariant under Weyl transformations.) If we assume that the worldvolume metric \(\gamma\) is nondegenerate and Lorentzian,
using the \(p+1\) diffeomorphisms we can go to the gauge
\begin{align}\label{eq:pbrane-gauge}
\gamma_{0 i}&=0,& \gamma_{00}&=- \det\left[\gamma_{i j}\right],
\end{align}
where $i,j=1,2,\ldots, p$. For M2-branes (\(p=2\)), this gauge was used in \cite{Hartnoll:2002th,Strickland-Constable:2021afa}; for particles (\(p=0\)), this is simply the proper-time gauge in which the einbein \(\sqrt{-\gamma_{00}}\) is gauge-fixed to unity \cite[Exercise~2.4]{Becker:2006dvp}.
Eliminating \(\gamma_{0i}\) and \(\gamma_{00}\) using \eqref{eq:pbrane-gauge}, one obtains
\begin{multline}
    S_p[\gamma_{ij},x] = \int_\Sigma\mathrm d^{p+1}z\,\Bigg(\frac12
        g_{\mu\nu}\left(\dot x^\mu\dot x^\nu
    -\det[\gamma_{ij}]\gamma^{ij}\partial_ix^\mu\partial_jx^\nu
    \right)\\
    +\frac12(p-1)\det[\gamma_{ij}]\tilde\Phi(x)^2+x^*A\Bigg).
\end{multline}
When \(p\ne1\), the equation of motion for \(\gamma_{ij}\) implies
\begin{equation}\label{eq:worldvolume-metric-eom}
    \gamma_{ij}=\tilde\Phi(x)^{-2}\partial_i x^\mu\partial_j x^\nu g_{\mu\nu},
\end{equation}
so that \(\gamma_{ij}\) may be eliminated from the action to obtain\footnote{Note that, although we assumed that the worldsheet metric $\gamma$ was non-degenerate to begin with, the rewritten action \eqref{eq:gauge-fixed-brane-action-components} is well-defined even for a degenerate $\gamma$.}
\begin{equation}\label{eq:gauge-fixed-brane-action-components}
    S_p[x] = \int_\Sigma\mathrm d^{p+1}z\,\left(\frac12
    g_{\mu\nu}\dot x^\mu\dot x^\nu
    -\frac12\Phi^2\lambda^m\lambda^ng^{(p)}_{mn}+A_{\mu m}\dot x^\mu\lambda^m\right),
\end{equation}
where
\begin{equation}
    \Phi(x) \coloneqq
    \tilde\Phi^{1-p}
    =\exp\mleft(-\frac{1-p}2\phi(x)\mright),
\end{equation}
we have redefined \(A_{\mu m}\) to absorb a factor of \(\frac{1}{p!}\), and we have defined 
\begin{equation}
   \lambda^m\equiv \lambda^{\mu_1\dotso\mu_p}\coloneqq\varepsilon^{i_1\dotso i_p}\partial_{i_1}x^{\mu_1}\dotsm\partial_{i_p}x^{\mu_p}
\end{equation}
introducing the index
\begin{equation}m \coloneqq\left[\mu_1 \ldots \mu_p\right]\end{equation}
for \(p\) totally antisymmetric tangent indices of \(M\), which are raised and lowered by
\begin{equation}
    g^{(p)}_{mn} \coloneqq\frac1{p!}g_{\mu_1\nu_1}\dotsm g_{\mu_p\nu_p},
\end{equation}
such that
\begin{equation}
\begin{aligned}
    \Phi^{2p}\det[\gamma_{ij}] &=\det\left[\partial_i x^\mu \partial_j x^\nu g_{\mu \nu}(x)\right]\\
    &=\frac1{p!}\varepsilon^{i_1\cdots i_p}\varepsilon^{j_1\cdots j_p}
    g_{\mu_1\nu_1}\partial_{i_1}x^{\mu_1}\partial_{j_1}x^{\nu_1}\dotsm
    g_{\mu_p\nu_p}\partial_{i_p}x^{\mu_p}\partial_{j_p}x^{\nu_p}\\
    &=g^{(p)}_{mn}\lambda^m\lambda^n.
\end{aligned}
\end{equation}
Note that when \(p=1\), even though \eqref{eq:worldvolume-metric-eom} is not a consequence of the equations of motion, using the Weyl symmetry one can still impose the conformal gauge
\begin{equation}
    \gamma_{\alpha\beta}=\begin{pmatrix}-1&0\\0&1\end{pmatrix},
\end{equation}
in which case one recovers \eqref{eq:gauge-fixed-brane-action-components} (and \(\Phi=1\) identically).

The equation of motion for $x^\mu$ in \eqref{eq:gauge-fixed-brane-action-components} generalises the geodesic equation we found for the particle \eqref{eq:particle-geodesic-equation} and takes the form
\begin{multline}\label{eq:equation-of-motion}
    \delta S = \int \left(\Phi\lambda^{\nu a} g^{(p-1)}_{ab}\nabla_\nu (\Phi\lambda^{\mu b})-\dot x^\nu\nabla_\nu \dot x^\mu+(\mathrm d A)^\mu{}_{\nu m}\dot x^\nu\lambda^m\right.\\\left.-\Phi\lambda^m g^{(p)}_{mn}\lambda^n\partial^\mu\Phi\right) g_{\mu\rho}\delta x^\rho,
\end{multline}
where we have expanded the $p$-multi index $m$ into a single index $\nu$ and a $(p-1)$-multi index $a$, and similarly $n=\mu b$.
During the gauge-fixing procedure we get first-order constraints.
For \(p=1\), we obtain the Virasoro condition for the vanishing of the stress--energy tensor.
For \(p=0\), we obtain the mass-shell condition
\begin{equation}\label{eq:mass-shell-condition}
    g_{\mu\nu}\dot x^\mu\dot x^\nu+\Phi^2(x)=0.
\end{equation}
For \(p>1\), these are the vanishing of the worldvolume total stress--energy tensor:
\begin{equation}
\begin{aligned}
    0&=T_{\alpha\beta}\\
    &=-2(-\det\gamma)^{-1/2}\frac{\delta\mathcal L}{\delta \gamma^{\alpha\beta}}\\
    &=g_{\mu\nu}\partial_\alpha x^\mu
    \partial_\beta x^\nu
    -2\gamma_{\alpha\beta}
    \left(
        \gamma^{\gamma\deltait}
         g_{\mu\nu}\partial_\gamma x^\mu
    \partial_\deltait x^\nu
    - (p-1)\tilde\Phi^2
    \right),
\end{aligned}
\end{equation}
where we have dropped some overall constant factors.
Since these are of first order, we can ignore them for the purposes of deriving conserved quantities.

\subsection{A Hamiltonian manifesting T/U\texorpdfstring{-}{‐}duality}\label{ssec:hamiltonian-duff-lu}
In this subsection we digress briefly to note that, by passing to generalised momenta in the sense of generalised geometry, we obtain a Hamiltonian that manifests T- and U-duality. It is well known that it is difficult to manifest T- or U-duality in the Lagrangian \cite{Percacci:1994aa,Duff:2015jka,Hu:2016vym} without resorting to double or exceptional field theory \cite{Sakatani:2016sko,Sakatani:2017vbd,Hatsuda:2024kuw}. For a related discussion of U-dual Hamiltonian formulations of branes, see \cite{Osten:2021fil}.
In this subsection (and this subsection only), we assume that \(g\) is invertible.

For $p=1$ (the fundamental string) and \(p=2\), \(d<5\) (the M2-brane) with \(\Phi=1\), it is possible to manifest \(\operatorname O(d,d)\) T-duality and U-duality by passing to the Hamiltonian picture using the Duff--Lu generalised $p$-brane metric $G_{\mathrm{DL}}$ \cite[(7.14)]{Duff:1990hn} (slightly generalised here to include a dilaton \(\Phi\))
\begin{equation}\label{eq:duff-lu-metric}
    G_{\mathrm{DL},MN}=\begin{pmatrix}
        g_{\mu\nu}+\Phi^{-2}A_{\mu m}A_\nu{}^m&\Phi^{-2}A_\mu{}^n\\
        \Phi^{-2}A_\nu{}^m&\Phi^{-2}g^{mn}
    \end{pmatrix}
\end{equation}
where \({}_M\coloneqq({}_\mu,{}^m)\), \({}_N\coloneqq({}_\nu,{}^n)\) are generalised indices. Its inverse is
\begin{equation}
    G_\mathrm{DL}^{MN}=\begin{pmatrix}
        g^{\mu\nu}&-A^\mu{}_n\\
        -A^\nu{}_m&\Phi^2g_{mn} + g^{\mu\nu}A_{\mu m} A_{\nu n}
    \end{pmatrix}.
\end{equation}
In particular, for \(p=0\), then \(G_{\mathrm{DL}}\) is nothing more than the Kaluza--Klein ansatz for the dimensional reduction of a \((d+1)\)-dimensional metric into a \(d\)-dimensional metric, graviphoton, and graviscalar \cite[(1.6)]{Bailin:1987jd}, also appearing in \cite{Hull:2007zu,Berman:2010is,Coimbra:2011ky}, and for \(d=3\) and \(4\), \(G_{\mathrm{DL}}\) reduces to the $\operatorname{SL}(3, \mathbb R)\times \operatorname{SL}(2; \mathbb R)$ and $\operatorname{SL}(5; \mathbb R)$ generalised metrics, respectively. 

From \eqref{eq:gauge-fixed-brane-action-components}, the canonical conjugate momentum to \(x^\mu\) is
\begin{equation}
    p_\mu = g_{\mu\nu}\dot x^\nu + A_{\mu m}\lambda^m,
\end{equation}
working with the convention that \(\epsilon^{01\dotso p}=1\).
String theory and supergravity can be naturally described \cite{Hull:2007zu,PiresPacheco:2008qik, Berman:2010is, Coimbra:2011nw, Coimbra:2011ky, Coimbra:2012af, Strickland-Constable:2013xta} in terms of (exceptional) generalised geometry \cite{Hitchin:2003cxu, Gualtieri:2003dx, Gualtieri:2007ng} (cf.\ \cite{Garcia-Fernandez:2020ope}). In the present case, in terms of the generalised tangent bundle \(\mathrm TM\oplus\bigwedge^p\mathrm T^*M\),\footnote{
    The generalised geometry of the \(p=0\) case, describing a particle coupled to a Maxwell field, is described in \cite[App.~C.1]{Strickland-Constable:2013xta}.
    The \(p=1\) case is the classical case \cite{Coimbra:2011nw,Garcia-Fernandez:2020ope} describing T-duality.
    The \(p=2\) case describes \(\operatorname{SL}(5;\mathbb R)\) U-duality \cite{Hull:2007zu,PiresPacheco:2008qik,Berman:2010is,Coimbra:2011ky}.
}
it is natural to introduce the generalised momentum
\begin{equation}
    P_M=(p_\mu,\lambda^m),
\end{equation}
where as above the generalised index \({}_M\coloneqq({}_\mu,{}^m)\) ranges over both cotangent and \(p\)-fold-antisymmetrised tangent directions.

Suppose that the worldvolume \(\Sigma\) is of the form \(\Sigma = \tilde\Sigma\times\mathbb R\) where the spatial coordinates \(z^1,\dotsc,z^p\) coordinatise \(\tilde\Sigma\) while the time coordinate \(z^0\) coordinatises \(\mathbb R\). Let the Lagrangian density be \(\mathcal L\).
Then the Hamiltonian is
\begin{equation}
\begin{aligned}
    H &= \int_{\tilde\Sigma}\mathrm d^pz\,
    p_\mu\dot x^\mu-
    \mathcal L\\
    &= \frac12\int_{\tilde\Sigma}\mathrm d^pz\,\left(g_{\mu\nu}\dot x^\mu\dot x^\nu
    +\Phi^2g^{mn}\lambda_m\lambda_n\right)\\
    &= \frac12\int_{\tilde\Sigma}\mathrm d^pz\,\left(g^{\mu\nu}p_\mu p_\nu
    +(\Phi^2 g_{mn} + g^{\mu\nu}A_{\mu m} A_{\nu n})\lambda^m\lambda^n
    -2g^{\mu\nu}A_{\mu m}p_\nu\lambda^m
    \right)\\
    &= \frac12\int_{\tilde\Sigma}\mathrm d^pz\,G_\mathrm{DL}^{MN}P_MP_N.
\end{aligned}
\end{equation}
The appearance of the Duff--Lu metric \eqref{eq:duff-lu-metric} shows that
this Hamiltonian formulation manifests the T- and U-dualities of the string or M2-brane.

This expression appears for the \(p=2\) case (M2-brane) in \cite[(4.26)]{Strickland-Constable:2021afa} and in \cite{Hatsuda:2012vm}. The Hamiltonian for arbitrary D$p$-branes (including all Neveu--Schwarz--Neveu--Schwarz and Ramond--Ramond fields) was presented in \cite{Hatsuda:2012uk}, and their expression reduces to ours when only the $(p+1)$-form potential is included. Similar expressions appear in \cite{Duff:1990hn,Berman:2010is,Sakatani:2016sko,Sakatani:2017vbd} in the context of double or exceptional field theory where one formally regards \(\lambda\) as the derivative of a new independent coordinate; we emphasise here that we have \emph{not} introduced any doubled coordinates.

\subsection{Symmetries of the \texorpdfstring{\(p\)-}{𝑝‐}brane worldvolume action}\label{ssec:brane-symmetries}
Given the form of the action \eqref{eq:gauge-fixed-brane-action-components}, which is quadratic in \(\dot x^\mu\) and \(\lambda^m\), it is natural to postulate transformations of the ansatz
\begin{equation}\label{eq:symmetry_ansatz}
    \delta x^\mu = K^\mu{}_{\nu_2\dotso\nu_k m_2\dotso m_l}v^{\nu_2}\dotsm v^{\nu_k}\lambda^{m_2}\dots\lambda^{m_l},
\end{equation}
where \(v^\mu\coloneqq\dot x^\mu\).
Now, there are two obvious classes of symmetries of this type: Killing vectors and time translations.
First, it is immediate that the action \eqref{eq:gauge-fixed-brane-action-components} is invariant under a transformation
\begin{equation}
    \delta x^\mu = \epsilon K^\mu
\end{equation}
with \(\epsilon\) an infinitesimal constant parameter and \(K\) a Killing vector such that
\begin{align}
    K\intprod \mathrm dA &= 0,&
    K\intprod \mathrm d \Phi &= 0.
\end{align}
Equivalently, using Cartan's magic formula, such a vector \(K\) satisfies
\begin{align}\label{eq:generalised-killing-vector}
    \mathcal L_K A &= \mathrm d\Lambda,&
    \mathcal L_K \Phi &= 0,
\end{align}
for some \(p\)-form \(\Lambda\); then \((K,\Lambda)\) may be regarded as a section of the generalised tangent bundle \(\mathrm TM\oplus\bigwedge^p\mathrm T^*M\). The first equation of \eqref{eq:generalised-killing-vector} can then be rephrased in terms of generalised geometry as saying that \((K,\Lambda)\) is a generalised Killing vector \cite{Grana:2008yw}. In the string theory case (\(p=1\)), this is equivalent to $K$ generating a $W$-symmetry, which in the presence of a Kalb--Ramond field is equivalent to demanding that $K$ generates a T-duality transformation \cite{Rocek:1991ps}. Similarly, for \(p=2\), generalised Killing vectors define symmetries for the M2-brane theory \cite{Strickland-Constable:2021afa}.

On the other hand, letting \(K^\mu{}_{\nu}\) be \(\delta^\mu_\nu\) in the ansatz \eqref{eq:symmetry_ansatz}, we obtain a transformation of the form
\begin{equation}
    \delta x^\mu = \epsilon \delta^\mu_\nu \dot x^\nu = \epsilon \dot x^\mu
\end{equation}
(again with \(\epsilon\) constant), which is simply an infinitesimal version of time translation\footnote{
    Of course, there is nothing special about the time direction that we have chosen; translation along any direction is a symmetry,
    but translation along other directions does not fall under the ansatz \eqref{eq:symmetry_ansatz}.
}
\begin{equation}
    x^\mu(t,z^i)\mapsto x^\mu(t+\Deltaup t,z^i).
\end{equation}
This is manifestly a symmetry of \eqref{eq:gauge-fixed-brane-action-components}.

Unfortunately, a direct but laborious computation shows that, when \(g\) is invertible and \(p\ge2\) and \(\Phi\ne0\), these are the only symmetries of the ansatz \eqref{eq:symmetry_ansatz}.

If \(g\) is non-invertible, however, there are additional symmetries. A general variation of \eqref{eq:equation-of-motion} can be rewritten as
\begin{multline}
    \delta S= -\int_\Sigma \delta x^\rho\Bigg(\Gamma_{\rho\mu\nu} v^\mu v^\nu
    -(\mathrm dA)_{\rho\mu m} v^\mu\lambda^m
    +\Gamma_{\rho mn}\lambda^m\lambda^n
    +g_{\rho\mu}\dot v^\mu\\
    -\Phi^2
    g_{\rho\sigma}
    \varepsilon^{i_1\dotso i_p}
    \partial_{i_1}\lambda^{\sigma\mu_2\dotso\mu_p}
    \partial_{i_2}x^{\nu_2}\dotsm\partial_{i_p}x^{\nu_p}
    g_{\mu_2\nu_2}\dotsm g_{\mu_p\nu_p}
    \Bigg),
\end{multline}
where \(\Gamma_{\rho\mu\nu}\) are the usual Christoffel symbols of the first kind, and
\begin{equation}
\begin{split}
    \Gamma_{\rho mn}=\Gamma_{\rho\underline{\mu a}\,\underline{\vphantom\mu\nu b}}&\coloneqq 
        \Phi(\partial_{[\rho|}\Phi)
        g^{(p-1)}_{ab}g_{|\mu]\nu}-\Phi^2g_{ab}^{(p-1)}\Gamma_{\rho\mu\nu}\\&~~~~~~-\Phi^2g_{\rho\nu}\Gamma_{a_2\mu b_2}g_{a_3b_3}\cdots g_{a_pb_p},
        \end{split}
\end{equation}
where the underline indicates that the index \(m\) has been expanded into \(\mu a=\mu a_2\dots a_p\) and \(n\) into \(\nu b=\nu b_2\dots b_p\).

It is then clear by examination that the transformation \eqref{eq:symmetry_ansatz} is a symmetry as long as the following conditions hold:
\begin{equation}\label{eq:degenerate_symmetries}
\begin{aligned}
    g_{\nu_1\mu}K^\mu{}_{\nu_2\dotso\nu_km_1\dotso m_l}&=0,&
    \Gamma_{\mu(\nu_0\nu_1}K^\mu{}_{\nu_2\dotso\nu_k)m_2\dotso m_k}&=0,\\
    (\mathrm dA)_{\mu(m_1|(\nu_1}K^\mu_{\nu_2\dotso\nu_k)|m_2\dotso m_k)}&=0,&
    \Gamma_{\mu(m_0m_1|}K^\mu{}_{\nu_2\dotso\nu_k|m_2\dotso m_l)}&=0.
\end{aligned}
\end{equation}
These expressions do not look covariant. However, when \(g\) is not invertible, \(g\) does not define an affine connection so it is unreasonable to expect covariant expressions.

\subsubsection{Symmetries of the tensionless \texorpdfstring{\(p\)-}{𝑝‐}brane worldvolume action}\label{ssec:tensionless}
The \(p\)-brane action \eqref{eq:gauge-fixed-brane-action-components} enjoys classical conformal symmetry in the tensionless \(\Phi\to0\) limit, where the action reduces to
\begin{equation}\label{eq:tensionless-action}
   S_p[x] = \int_\Sigma\mathrm d^{p+1}z\,\left(
        \frac12g_{\mu\nu}\dot x^\mu\dot x^\nu +  A_{\mu m}\dot x^\mu
     \lambda^m\right).
\end{equation}
Let us repeat the analysis of \cref{ssec:brane-symmetries} for this tensionless limit. (The tensionless limit requires first-class constraints \cite{Lindstrom:2000pp}, which we again ignore for the purpose of constructing conserved quantities.)

Let \(K\) be a totally symmetric tensor of rank \(k\) on spacetime \(M\), and suppose first that the metric $g$ is nondegenerate. Consider the transformation
\begin{equation}
    \delta x^\mu = \epsilon K^\mu{}_{\nu_2\dotso\nu_k}v^{\nu_2}\dotsm v^{\nu_k}
\end{equation}
where \(\epsilon\) is an infinitesimal constant (independent of the worldvolume coordinates). The corresponding variation of the action is given by
\begin{equation}
    \delta S=\delta_{(k+1,0)}S+\delta_{(k,1)}S
\end{equation}
where
\begin{equation}
\begin{aligned}
    \delta_{(k+1,0)}S&\coloneqq-\int\Gamma^\rho_{\mu_0\mu_1}K_{\rho\mu_2\dotso\mu_k}v^{\mu_0}\dotsm v^{\mu_2}-\int \dot v^{\mu_1}K_{\mu_1\dotso\mu_k}v^{\mu_2}\dotsm v^{\mu_k}\\
    &= \frac1k\int(\nabla_{\mu_0}K_{\mu_1\dotso\mu_k})v^{\mu_0}\dotsm v^{\mu_k},\\
    \delta_{(k,1)}S&\coloneqq \int(\mathrm dA)^\rho_{\mu_1 m}K_{\rho\mu_2\dotso \mu_k}v^{\mu_1}\dotsm v^{\mu_k}\lambda^m,
\end{aligned}
\end{equation}
where we have performed some integrations by parts.

Thus, cancellation of all terms in \(\delta S\) requires
\begin{equation}\label{eq:generalised-conformal-killing-tensor}
\begin{aligned}
    0 &= \nabla_{(\mu_0}K_{\mu_1\dotso\mu_k)},&
    0 &= g^{\mu_0\nu_1}(\mathrm dA)_{\mu_0(\mu_1\vert\mu_2\dotso\mu_{p+1}}K_{\nu_1\vert\nu_2\dotso\nu_k)}.
\end{aligned}
\end{equation}
corresponding to the vanishing of \(\delta_{(k+1,0)}S\) and \(\delta_{(k,1)}\) respectively.
Unlike the tensile case, in the tensionless case we obtain Killing tensors of all ranks even with a nondegenerate metric.

In addition, of course, when the metric \(g\) is degenerate, any transformation \eqref{eq:symmetry_ansatz} satisfying \eqref{eq:degenerate_symmetries} is still a symmetry; in the tensionless case \(\Gamma_{\mu m_0m_1}=0\), so that the last equation of \eqref{eq:degenerate_symmetries} holds trivially.

\subsection{Symmetries of the charged particle with position\texorpdfstring{-}{‐}dependent mass}\label{ssec:particle-symmetry}
For \(p=0\), the action \eqref{eq:gauge-fixed-brane-action-components} reduces to
\begin{equation}\label{eq:particle-action-general}
    S[x] = \int_\Sigma\mathrm dz\,\left(
        \frac12g_{\mu\nu}\dot x^\mu\dot x^\nu
        -\frac12\Phi(x)^2+A_\mu\dot x^\mu
    \right)
\end{equation}
which describes a scalar particle with worldline \(\Sigma\) of (in general position-dependent) mass \(\Phi(x)\) living in a spacetime \(M\) with (pseudo-)Riemannian metric \(g\) and coupled to a Maxwell field \(A\) with unit charge.
The Duff--Lu metric \eqref{eq:duff-lu-metric} in this case is simply the Kaluza--Klein metric in \(d+1\) dimensions. This reflects the fact that a free particle of constant mass on \(M\times\mathbb S^1\) dimensionally reduces to a Kaluza--Klein tower of scalar fields \(\phi_n\) labelled by an integer \(n\) which do not interact with each other and thus can be considered individually; then for \(n\ne0\) the scalar field \(\phi_n\) is coupled to the graviphoton and has a position-dependent mass given by the graviscalar.
In the case \(p=0\), the action \eqref{eq:gauge-fixed-brane-action-components} enjoys additional symmetries beyond that described in \cref{ssec:brane-symmetries} essentially because \(\lambda=1\) is constant so that would-be terms depending on the derivatives of \(\lambda\) automatically vanish.

Let \(\lbrace K^{(k)} \rbrace\) (with \(k\in\{1,2,\dotsc\}\)) be a collection of tensor fields of rank \((1,k-1)\) on the target space \(M\).
The ansatz \eqref{eq:symmetry_ansatz} simplifies since now \(\lambda=1\); thus we write a slight generalisation of this ansatz that includes the whole collection of tensors
\begin{equation}\label{eq:particle-variation-variable-mass}
    \delta x^\mu = \epsilon\sum_{k=1}^\infty K^{(k)\mu}{}_{\nu_2\dotso\nu_k} v^{\nu_2}\dotsm v^{\nu_k},
\end{equation}
where \(\epsilon\) is the infinitesimal symmetry parameter constant along the worldline.

Any transformation \eqref{eq:symmetry_ansatz} satisfying \eqref{eq:degenerate_symmetries} is still a symmetry.
In addition, if we further assume that \(g_{\mu\nu_1}K^{(k)\mu}{}_{\nu_2\dotso\nu_k}\) is totally symmetric for each $K^{(k)}$, then one can integrate by parts so that the variation of the action \eqref{eq:particle-action-general} under \eqref{eq:particle-variation-variable-mass} is
\begin{equation}
\begin{split}
    \delta S = \sum_{k=-1}^\infty\epsilon\int_\Sigma \left(\frac1k\nabla_{\mu_0}K^{(k)}_{\mu_1\dotso\mu_k}
    \right.+&\left.(\mathrm dA)_{\nu\mu_0} K^{(k+1)\nu}{}_{\mu_1\dotso\mu_k}\right.
    \\-&\left.K^{(k+2)\nu}{}_{\mu_0\dotso\mu_k}\Phi\partial_\nu\Phi\right)v^{\mu_0}\dotsm v^{\mu_k},
    \end{split}
\end{equation}
where terms depending on \(k^{-1}K^{(k)}\) are understood to be identically zero when \(k\le0\),
and where (when \(g\) is degenerate) the first term should be understood in the same sense as the last line of \eqref{eq:degenerate_covariant_derivative}.
Thus a symmetry of the particle worldline action is given by a sequence of totally symmetric tensors \(K^{(1)},K^{(2)},\dotsc\) such that 
\begin{equation}\label{eq:particle_generalised_killing}
    X^{(k+1)}_{\mu_0\dotso\mu_k}\coloneqq k^{-1}\nabla^{\vphantom(}_{(\mu_0}K^{(k)}_{\mu_1\dotso\mu_k)}+(\mathrm dA)^{\vphantom(}_{\nu(\mu_0} K^{(k+1)\nu}_{\mu_1\dotso\mu_k)}
    -K^{(k+2)\nu}{}_{\mu_0\dotso\mu_k}\Phi\partial_\nu\Phi
    =0,
\end{equation}
where again the first term should be understood in the same sense as \eqref{eq:degenerate_covariant_derivative} when \(g\) is degenerate.
This equation involving \(\mathrm dA\) may be seen as a symmetrised version of the \emph{quasi-instanton condition} found for gauge fields in two-dimensional sigma-models in \cite{deLaOssa:2018iij}.

However, this is not the most general symmetry since we did not use yet the mass-shell constraint \eqref{eq:mass-shell-condition}, using which we can extract traces out of \(X^{(k)}\); this will yield additional \emph{trivial symmetries}, whose associated conserved quantities vanish, cf.\ \cite{MartinezAlonso:1979fej}, \cite[p.~377]{zbMATH00478435}, \cite[§3.1.5]{Henneaux:1994lbw}.
For simplicity, assuming \(g\) to be invertible, we can decompose \(X^{(k)}\) into irreducible representations of the spacetime Lorentz group \(\operatorname O(g)\), i.e.\ as a sum of totally traceless totally symmetric tensors \(X^{(k,i)}_{\mu_1\dotso\mu_{k-2i}}\) for \(i\in\{0,\dotsc,\lfloor k/2\rfloor\}\):
\begin{equation}
\begin{aligned}
    X^{(k)}_{\mu_1\dotso\mu_k}
    &\eqqcolon\sum_{i=0}^{\lfloor k/2\rfloor}
    g^{\vphantom(}_{(\mu_1\mu_2}\dotsm g^{\vphantom(}_{\mu_{2i-1}\mu_{2i}\vphantom(}
    X^{(k,i)}_{\mu_{2i+1}\dotso\mu_k)}\\
    &=
    X^{(k,0)}_{\mu_1\dotso\mu_k}
    +
    g^{\vphantom(}_{(\mu_1\mu_2}
    X^{(k,1)}_{\mu_3\dotso\mu_k)}
    +
    g^{\vphantom(}_{(\mu_1\mu_2}g^{\vphantom(}_{\mu_3\mu_4\vphantom(}
    X^{(k,2)}_{\mu_3\dotso\mu_k)}
    +\dotsb.
\end{aligned}
\end{equation}
Then it is clear that the most general symmetry condition is
\begin{equation}\label{eq:particle-cond-general}
    \sum_{i=0}^\infty
    (-\Phi^2)^i
    X^{(k+2i,i)}_{\mu_1\dotso\mu_k}
    = 0.
\end{equation}

Note that if we have \(K^{(k)}=0\) for \(k \ne 1\) (\(k=1\) corresponds to the case of a vector field), the condition \eqref{eq:particle-cond-general} reduces to
\begin{align}
    \mathcal L_Kg&=0,&\mathcal L_KA&=\mathrm d\zeta, &K\Phi&=0,
\end{align}
which is that of a generalised Killing vector for the generalised geometry $TM\oplus\mathbb{R}\times M$ \cite[App.~C.1]{Strickland-Constable:2013xta} that further preserves the dilaton.
More generally, if \(K^{(i)}=0\) for \(i\ne k\), then
the condition \eqref{eq:particle-cond-general} reduces to
\begin{align}
    \nabla_{(\mu_0}K^{(k)}_{\mu_1\dotso\mu_k)}&=0,&(\mathrm dA)^{\mu\nu_1}K_{\mu_1\dotso\mu_k}&=0,&
    (\partial^{\mu_1}\Phi)K_{\mu_1\dotso\mu_k}&=0,
\end{align}
that is, \(K^{(k)}\) is a Killing tensor that preserves the field strength \(\mathrm dA\) and the dilaton.

In the massless case \(\Phi=0\), then \eqref{eq:particle-cond-general} reduces to the condition that 
\begin{equation}
    k^{-1}\nabla_{(\mu_0}K^{(k)}_{\mu_1\dotso\mu_k)}+(\mathrm dA)^\rho{}_{(\mu_0} K^{(k+1)}_{\mu_1\dotso\mu_k)\rho}=g_{(\mu_0\mu_1}\tilde K_{\mu_2\dotso\mu_k)}
\end{equation}
for some totally symmetric tensors \(\tilde K_{\mu_2\dotso\mu_k}\); when \(\mathrm dA=0\), this simply means that each \(K^{(k)}\) must be a conformal Killing tensor.

\subsection{Symmetries of the bosonic string}\label{ssec:howe-papadopoulos-string}
For \(p=1\), the action \eqref{eq:brane-action} describes a bosonic string moving in a spacetime \(M\) coupled to the metric \(g\) and the Kalb--Ramond field \(A\) (which, following tradition, we write as \(B\) in this section).
There is no coupling to the dilaton \(\phi\), but if one wishes one can add an Euler-class term to which the dilaton couples.

The worldsheet Polyakov action of a bosonic string coupled to NS--NS background fields \((g,B,\Phi)\) is
\begin{equation}
    S[x,\gamma] = \int_\Sigma\mathrm d^2z\,\sqrt{|\det \gamma|}\frac12g(x)_{\mu\nu}\gamma^{\alpha\beta}\partial_\alpha x^\mu\partial_\beta x^\nu+x^*B+\phi(x)R,
\end{equation}
where \(x\colon\Sigma\to M\) is the embedding map,
\(\gamma\) is the worldsheet metric, \(x^*B\) is the pullback of the two-form \(B\) onto the worldsheet, and \(R\in\Omega^2(\Sigma)\) is the curvature two-form (i.e.\ the Euler form) of \(\gamma\) normalised so that the Euler characteristic \(\operatorname\chiup(\Sigma)\) of \(\Sigma\) is given by
\begin{equation}
    \operatorname\chiup(\Sigma) = \int_\Sigma R.
\end{equation}
This action is invariant under Weyl transformations as well as two-dimensional diffeomorphisms; using these three degrees of gauge freedom, we may gauge-fix the three independent components of \(\gamma\) to go to the conformal gauge where \(\gamma\) is the Minkowski metric, which in lightcone coordinates \(z=(z^+,z^-)\) reads
\begin{align}
    \gamma_{+-}=\gamma_{-+}&=1,&\gamma_{++}=\gamma_{--}&=0,
\end{align}
so that the action simplifies to
\begin{equation}
    S[x] = \int\mathrm d^2z\,N(x)_{\mu\nu}\partial_+ x^\mu\partial_- x^\nu
    + \Phi(x)R,
\end{equation}
where
\begin{equation}
    N_{\mu\nu}\coloneqq g_{\mu\nu}+B_{\mu\nu},
\end{equation}
at the cost of imposing by hand the Virasoro constraint that the stress--energy must vanish:
\begin{equation}\label{eq:virasoro-constraint}
    0 = g_{\mu\nu}\partial_+ x^\mu\partial_+ x^\nu = g_{\mu\nu}\partial_- x^\mu\partial_- x^\nu.
\end{equation}
We define
\begin{align}
    v_\pm^\mu &\coloneqq \partial_\pm x^\mu,&
    v_{+-}^\mu&\coloneqq\partial_\pm v_\mp^\mu = \partial_+\partial_-x^\mu.
\end{align}
A general variation of the action is
\begin{equation}
    \delta S = \int \left(-2g_{\mu\nu}v_{+-}^\nu 
    +
    (\partial_\mu N_{\nu\rho})v_+^\nu v_-^\rho
    +
    (\partial_\mu\Phi)R\right)\delta x^\mu.
\end{equation}

Let \(K\) be a totally symmetric tensor of rank $k$ on spacetime \(M\), and suppose first that the metric $g$ is nondegenerate.
We now postulate the transformation
\begin{equation}\label{eq:string-howe-papadopoulos-variation}
    \delta x^\nu(z) = \epsilon(z^+)K^{\nu}{}_{\mu_2\dotso\mu_k}v_+^{\mu_2}\dotso v_+^{\mu_k}.
\end{equation}
Then
\begin{equation}
    \delta S= \delta_1S+\delta_2S+\delta_3S
\end{equation}
where up to total derivatives 
\begin{equation}
\begin{aligned}
    -\frac12\delta_1S &= \int \epsilon K_{\mu_1\dotso\mu_k}(\partial_-v_+^{\mu_1})v_+^{\mu_2}\dotsm v_+^{\mu_k}\\
    &=\frac1k\int \epsilon K_{\mu_1\dotso\mu_k}\partial_-(v_+^{\mu_1}\dotsm v_+^{\mu_k})\\
    &=-\frac1k\int \epsilon \partial_\nu K_{\mu_1\dotso\mu_k}v_-^\nu v_+^{\mu_1}\dotsm v_+^{\mu_k},\\
    \delta_2S &= \int\epsilon(\partial_{\mu_1} N_{\nu\rho})v_+^\nu v_-^\rho K^{\mu_1}{}_{\mu_2\dotso\mu_k}v_+^{\mu_2}\dotsm v_+^{\mu_k},\\
    \delta_3 S &= \int\epsilon R(\partial_{\mu_1}\phi)K^{\mu_1}{}_{\mu_2\dotso\mu_k}v_+^{\mu_2}\dotsm v_+^{\mu_k}.
\end{aligned}
\end{equation}
The first two terms combine into
\begin{equation}
    \delta_1S+\delta_2S = -\frac1p\int\epsilon(\nabla^+_\nu K_{\mu_1\dotso\mu_k})v_-^\nu v_+^{\mu_1}\dotsm v_+^{\mu_k},
\end{equation}
where we denote by \(\nabla^\pm\) the affine connections on \(\mathrm TM\) with torsion given by \(\pm\mathrm dB\) respectively. Hence a sufficient condition for this to be a symmetry is
\begin{align}\label{eq:condition-string-naive}
    \nabla^+_\nu K_{\mu_1\dotso\mu_k} &= 0,&
    (\partial_{\mu_1}\phi)K^{\mu_1}{}_{\mu_2\dotso\mu_k}&=0,
\end{align}
that is, \(K\) must be covariantly constant with respect to the connection \(\nabla^+\). Of course, by working with \(v_-\) rather than \(v_+\), one obtains the same conditions as \eqref{eq:condition-string-naive} except with the connection \(\nabla^-\) rather than \(\nabla^+\).

We have not, however, used the Virasoro constraint \eqref{eq:virasoro-constraint}. Using the constraint we can relax the symmetry to totally symmetric tensors whose totally traceless components are covariantly constant:
\begin{align}
    \nabla^+_\nu\tilde K_{\mu_1\dotso\mu_k} &= 
    0,&
    (\partial_{\mu_1}\phi)\tilde K^{\mu_1}{}_{\mu_2\dotso\mu_k}&=0,
\end{align}
where \(\tilde K\) denotes the totally traceless component of \(K\). By a similar argument, transformations involving both \(v_+\) and \(v_-\) are also symmetries whenever a certain trace (involving either two \(+\) or two \(-\) indices) vanishes. Thus, the general symmetry transformation is a linear combination of the following four kinds:
\begin{subequations}
\begin{equation}
    \delta x^\rho
    = \epsilon(z^+)g^{\rho\mu_1}\tilde K_{\mu_1\dotso\mu_k}v_+^{\mu_1}\dotsm v_+^{\mu_k},\qquad
    \nabla^+_\nu \tilde K_{\mu_1\dotso\mu_k} = 0 = (\partial^\mu\phi)\tilde K_{\mu_1\dotso\mu_k}
\end{equation}
with \(\tilde K\) totally symmetric and totally traceless;
\begin{equation}
    \delta x^\rho
    = \epsilon(z^-)g^{\rho\mu_1}\tilde K_{\mu_1\dotso\mu_k}v_-^{\mu_1}\dotsm v_-^{\mu_k},\qquad
    \nabla^-_\nu \tilde K_{\mu_1\dotso\mu_k} = 0 = (\partial^\mu\phi)\tilde K_{\mu_1\dotso\mu_k}
\end{equation}
with \(\tilde K\) totally symmetric and totally traceless; and
\begin{equation}
    \delta x^\rho
    = \epsilon(z^+,z^-)g_{(\mu_1\mu_2}K^\rho{}_{\mu_3\dotso\mu_k)\nu_1\dotso\nu_l}v_+^{\mu_1}\dotsm v_+^{\mu_k}v_-^{\nu_1}\dotso v_-^{\nu_l},
\end{equation}
with \(K^\rho{}_{\mu_3\dotso\mu_k\nu_1\dotso\nu_l}=K^\rho{}_{(\mu_3\dotso\mu_k)(\nu_1\dotso\nu_l)}\); and
\begin{equation}
    \delta x^\rho
    = \epsilon(z^+,z^-)g_{(\nu_1\nu_2|}K^\rho{}_{\mu_1\dotso\mu_k|\nu_3\dotso\nu_l)}v_+^{\mu_1}\dotsm v_+^{\mu_k}v_-^{\nu_1}\dotso v_-^{\nu_l},
\end{equation}
with \(K^\rho{}_{\mu_1\dotso\mu_k\nu_3\dotso\nu_l}=K^\rho{}_{(\mu_1\dotso\mu_k)(\nu_3\dotso\nu_l)}\).
\end{subequations}

The first two kinds of symmetries correspond to holomorphic and antiholomorphic currents, respectively, if they are not anomalous. The third and fourth kinds are \emph{trivial symmetries}, whose associated conserved quantities vanish, cf.\ \cite{MartinezAlonso:1979fej}, \cite[p.~377]{zbMATH00478435}, \cite[§3.1.5]{Henneaux:1994lbw}.

In addition, of course, if \(g\) is non-invertible, then any transformation \eqref{eq:symmetry_ansatz} satisfying \eqref{eq:degenerate_symmetries} is still a symmetry, and in the tensionless case, any transformation \eqref{eq:symmetry_ansatz} satisfying \eqref{eq:generalised-conformal-killing-tensor} is a symmetry, generalising the results of \cite{Lindstrom:2022iec} to the case with a Kalb--Ramond field.

\section{Manifestations of brane symmetries}\label{sec:applications}
As seen in \cref{sec:warmup}, the symmetries of a brane action are avatars not only of the invariants of the generalised geodesics traced out by the branes but also of the symmetries of the bulk theory.
We briefly examine these for the symmetries of tensile and tensionless branes.

\subsection{Worldvolume Noether charges as invariants of geodesic motion}
\label{sec:noether-charge}
The trajectories swept out by worldvolumes of branes can be often regarded as geodesics in a generalised-geometric sense \cite{Strickland-Constable:2021afa}.
Just as for the particle case in \cref{sec:warmup}, the symmetries provide invariants of these generalised geodesics \cite{Prochazka:2024xyd}.
If a transformation  \eqref{eq:symmetry_ansatz}
is a symmetry of a brane action, by Noether's theorem,
the conserved quantities corresponding to them are \(j^\alpha_\mu K^\mu{}_{\nu_2\dotso\nu_km_2\dotso m_l}v^{\nu_2}\dotsm v^{\nu_k}\lambda^{m_2}\dotsm\lambda^{m_l}-J^\alpha\) where
\begin{equation}
    j^\alpha_\mu \coloneqq \frac{\delta \mathcal L}{\delta\partial_\alpha x^\mu},\qquad \delta_\epsilon \mathcal L =\epsilon \partial_\alpha J^\alpha
\end{equation}
 so that
\begin{align}
    j^0_\mu &= g_{\mu\nu}\dot x^\nu+A_{\mu m}\lambda^m=P_\mu,\\
    j^i_\mu &= (g_{mn}\lambda^m+A_{\nu m}\dot x^\nu)\frac{\delta\lambda^m}{\delta\partial_ix^\mu},
\end{align}
and Noether's theorem then states that
\begin{equation}
    \partial_\alpha\left(j^\alpha_\mu K^\mu{}_{\nu_2\dotso\nu_km_2\dotso m_l}v^{\nu_2}\dotsm v^{\nu_k}\lambda^{m_2}\dotsm\lambda^{m_l}-J^\alpha\right)=0.
\end{equation}
In particular, for a compact boundaryless \(\tilde\Sigma\) with \(\Sigma=\tilde\Sigma\times\mathbb R\), 
\begin{equation}
    \frac{\mathrm d}{\mathrm dt}\int_{\tilde\Sigma}\left(j^0_\mu K^\mu{}_{\nu_2\dotso\nu_km_2\dotso m_l}-J^0\right)
    = 0.
\end{equation}
These can be seen as brane generalisations of the statements for geodesics given in \cref{eq:killing_geodesic_particle,eq:conformal_killing_geodesic_particle}.

As an illustrative example, let us consider the geodesic invariants of the string. 
As explained in \cref{ssec:howe-papadopoulos-string},
the classical conformal symmetries of the string mean that we have additional conserved quantities.
The equation of the motion for the bosonic string is
\begin{equation}
    \partial_+ (g_{\mu\nu}\partial_-x^\mu)=0=\partial_- (g_{\mu\nu}\partial_+x^\mu).
\end{equation}
(The fact that we have two rather than one equations is because in addition to the usual equation of motion we have the compatibility condition between the coordinates, as explained in \cite[(2.30)]{Strickland-Constable:2021afa}.)

Now, if \(K\) is a totally symmetric tensor of rank $k$ covariantly constant with respect to \(\nabla^+\), then the following quantity
\begin{equation}
    Q^+_K\coloneqq K_{\mu_1\dotso\mu_k}\partial_+x^{\mu_1}\dotsm\partial_+x^{\mu_k}
\end{equation}
is conserved along the worldsheet in the \(z^-\) direction in the following sense:
\begin{equation}
\begin{aligned}
    \partial_-x^\mu\nabla_\mu Q^+_K&=\partial_-x^\mu\nabla_\mu \left(K_{\mu_1\dotso\mu_k}\partial_+x^{\mu_1}\dotsm\partial_+x^{\mu_k}\right)
    \\
    &=\partial_-x^\mu(\nabla_\mu K_{\mu_1\dotso\mu_k})\partial_+x^{\mu_1}\dotsm\partial_+x^{\mu_k}\\&\qquad+ k K_{\mu_1\dotso\mu_k}(\partial_-x^\mu\partial_+x^{\mu_1})\partial_+x^{\mu_2}\dotsm\partial_+x^{\mu_k}=0.
\end{aligned}
\end{equation}
More generally, \(K\) multiplied by any function of \(z^-\) will be conserved similarly.

Similar statements of course hold for totally symmetric tensors covariantly constant with respect to \(\nabla^-\).

\subsection{Brane analogues of W\texorpdfstring{-}{‐}algebras}
\label{sec:w-algebra}
The \((p+1)\)-dimensional field theory \eqref{eq:brane-action} on the worldvolume \(\Sigma\) may be first-quantised to yield a quantum-mechanical system describing a brane living on \(M\).

In the Hamiltonian quantum formalism, the quantities \(x^\mu\) and \(v^\mu\) and \(\lambda^m\) are promoted to operators.
One can then define the operator
\begin{equation}
    J=K^\mu{}_{\nu_2\dotso\nu_km_2\dotso m_l}(x)v^{\nu_2}\dotsm v^{\nu_k}\lambda^{m_1}\dotsm\lambda^{m_l}
    +\mathcal O(\hbar),
\end{equation}
where there is an \(\mathcal O(\hbar)\) ambiguity due to ordering issues. One then has
\begin{equation}
    [H,J]=\mathcal O(\hbar)
\end{equation}
where \(H\) is the Hamiltonian.
If one can choose the operator ordering in \(J\)  such  that \([H,J]=0\), then by definition the classical symmetry is not anomalous and can be promoted to a symmetry of the quantised system, assuming that a quantisation is possible.

\paragraph{Particle analogue of W-algebra}
This is particularly concrete in the case \(p=0\) of a particle. For simplicity suppose that there is constant mass \(m\) and that there is no Maxwell field. Let \(M=\tilde M\times\mathbb R\) and \(\Sigma=\tilde\Sigma\times\mathbb R\) be the space--time decomposition. Solving the on-shell constraint yields the Hilbert space \(\operatorname L^2(\tilde M;\mathbb C)\). Let \(i,j,\dotsc\in\{1,\dotsc,d\}\) be tangent indices for \(\tilde M\).

On this, we have the operators \(x^i\) and the conjugate momentum
\begin{equation}P_i=-\mathrm i\hbar\nabla_i + \mathcal O(\hbar^2),\end{equation}
as well as
\begin{align}
    x^0 &=t,&P_0 &=\sqrt{m^2+\hbar^2\dot x^i\dot x^jg_{ij}(x)} + \mathcal O(\hbar^2),
\end{align}
where the term \(\mathcal O(\hbar^2)\) reflects ordering ambiguities.

Then corresponding to each Killing tensor \(K\), we have the operator
\begin{equation}
    J = \frac1{p!}K^{\mu_1\dotso\mu_k}P_{\mu_1}\dotsm P_{\mu_k}(1+\mathcal O(\hbar)).
\end{equation}
This quantity is then conserved by time evolution up to \(\mathcal O(\hbar)\) terms from ordering issues.
If the operators \(P_\mu\) and \(J\) can be ordered appropriately such that \(J\) is exactly conserved, then the corresponding symmetry is not anomalous; if this is not possible, then the corresponding symmetry is anomalous. 

\paragraph{String W-algebra}
In the string case, conformal symmetry enhances the W-algebra. For totally symmetric tensors that are covariantly constant with respect to \(\nabla^+\) or \(\nabla^-\), one obtains holomorphic or antiholomorphic currents, so that if anomalies are absent one obtains an infinite-dimensional algebra known as the W-algebra (reviewed in \cite{Bouwknegt:1992wg,Dickey:1997wia,Prochazka:2024xyd}). For example, the stress--energy tensor \(T=T(z)+\bar T(\bar z)\) of conformal weights \((2,0)\) and \((0,2)\) corresponds to the Killing tensor \(g\) which always exists and is hence always part of the W-algebra. A covariantly constant totally symmetric tensor of rank \(k\) corresponds to a W-algebra generator of weight \(\frac{k}{2}\) \cite{Hull:1985jv}.

\subsection{Conserved quantities for the bulk spacetime theory}
\label{sec:conserved-quantities}
Some (but not all) of the brane theories \eqref{eq:brane-action} can be third-quantised to form bulk theories.
When \(p=0\), one obtains a free scalar field theory.
(Scalar because we do not include  spin degrees of freedom, and free because the action is only defined on smooth one-manifolds rather than on arbitrary graphs).
When \(p=1\) and \(d=26\), one obtains a string field theory with an infinite tower of fields and a tachyon, which is not a problem for computing scattering amplitudes and solutions to the classical equations of motion.
When \(p=2\) and \(d=4\), this is expected to describe certain compactifications of M-theory on an \(M_4\times M_7\).

\paragraph{Particle case: free field theory.}
A system of particles living in a \(d\)-dimensional target space can be described in one of two equivalent ways: in terms of a theory on the one-dimensional worldline \(\Sigma\), and in terms of a second-quantised (or third-quantised) description of a field theory in \(d\) dimensions on the target space \(M\). If one only considers smooth \(\Sigma\), then this does not fix interaction terms of the target-space theory on \(M\); for this one must allow for worldline actions where \(\Sigma\) is a graph rather than a mere one-manifold. That is, a worldline action of particles defined on smooth one-manifolds does not know how to interact (unlike the analogous situation for strings).
Since we only consider actions on worldlines without singularities,
the symmetries found in \cref{ssec:particle-symmetry} translate to the symmetries of a system of free scalar particles with position-dependent masses on a manifold \(M\) coupled to a background Maxwell field \(A\).

Concretely, on a smooth manifold \(M\), consider a Maxwell field \(A\) and a metric \(g\) as well as a graviscalar \(\Phi(x)\). Then we have the operator
\begin{equation}
    \square_A = g^{\mu\nu}(\nabla-\mathrm iA)_\mu(\nabla-\mathrm iA)_\nu-\Phi(x)^2
\end{equation}
such that \(\mathcal C\coloneqq \ker\square_A\) is the space of solutions to the equation of motion for a free complex scalar field with spatially varying mass \(\Phi(x)\) coupled to the Maxwell field and the metric. The space of solutions to the equations of motion is
\begin{equation}
    \mathcal C = \ker\square_A.
\end{equation}
Symmetries of the bulk theory in the present context are those differential operators \(T\) such that
\begin{equation}
    \square_AT=T'\square_A
\end{equation}
for some differential operator \(T'\). 

Now, \(\square_A\) may be obtained as the dimensional reduction of a d'Alembertian operator in one higher dimension.
Thus, the theorem in \cite{Eastwood:2002su} cited above in \cref{sec:warmup} may be dimensionally reduced to show that indeed the symmetries of the bulk theory (in the above sense) are in bijection with symmetries of the worldline theory in \cref{ssec:particle-symmetry}.

\paragraph{Symmetries of string field theory and M-theory.}
From the bulk point of view, bosonic string theory corresponds to a 26-dimensional interacting field theory of \(\mathcal N=0\) supergravity coupled to a scalar tachyon and an infinite tower of massive stringy fields.
Unlike (the smooth-worldline action for) a particle, (the smooth-worldsheet action for) a string \emph{does} know how to interact via the sum-over-topologies, and interacting theories have fewer symmetries than free theories.

From this perspective the fact that the worldsheet action for tensile strings moving in a target space with nondegenerate metric lacks symmetries apart from worldsheet coordinate translations, target-space Killing vectors, and W-symmetries from covariantly constant tensors reflects the well known swampland conjecture that quantum gravity lacks   global symmetries \cite{Banks:2010zn}.
On the other hand, the presence of more interesting symmetries for tensionless strings or in the presence of degenerate metrics may suggest that any target-space quantisation may be more trivial so as to evade the no-global-symmetries conjecture.

For \(p=2\) and with \(\Phi=1\) constant, the action \eqref{eq:brane-action} describes the bosonic sector of the M2-brane coupled to a background metric \(g\) and the supergravity three-form potential \(A_{\mu\nu\rho}\) \cite{Bergshoeff:1987qx}.
Since we are working with bosonic objects only, we consider only the bosonic sector of M-theory,
which has been argued to be governed by the very extended Kac--Moody algebra \(\operatorname D_8^{+++}\) \cite{Glennon:2024ict,Glennon:2025stv}. Again, the fact that for two-branes we do not find any interesting symmetries is consistent with the no-global-symmetries conjecture for quantum gravity.

\section*{Acknowledgements}
The authors thank Charles Strickland-Constable\textsuperscript{\orcidlink{0000-0003-0294-1253}} for helpful discussion and pointers to the literature. Hyungrok Kim thanks Guillermo Arias-Tamargo\textsuperscript{\orcidlink{0000-0002-0713-789X}} for helpful discussion.
Leron Borsten is grateful for the hospitality of the Theoretical Physics group,
Blackett Laboratory, Imperial College London.
Mateo Galdeano is supported by an EPSRC New Investigator Award, grant number EP/X014959/1. No new data was collected or generated during the course of this research.

\newcommand\cyrillic[1]{\fontfamily{Domitian-TOsF}\selectfont \foreignlanguage{russian}{#1}}

\bibliographystyle{unsrturl}
\bibliography{biblio}
\end{document}